# Predicting Coordinated Actuated Traffic Signal Change Times using LSTM Neural Networks

**Seifeldeen Eteifa**
Charles E. Via, Jr. Department of Civil and Environmental Engineering
Virginia Tech, Blacksburg, Virginia, 24061
Email: seteifa@vt.edu

**Hesham A Rakha (Corresponding Author)**
Samuel Reynolds Pritchard Professor of Engineering
Charles E. Via, Jr. Department of Civil and Environmental Engineering
Virginia Tech, Blacksburg, Virginia, 24061
Email: hrakha@vt.edu
ORCiD: https://orcid.org/0000-0002-5845-2929

**Hoda Eldardiry**
Associate Professor, Computer Science Department
Virginia Tech, Blacksburg, Virginia, 24061
Email: hdardiry@vt.edu

## ABSTRACT
Vehicle acceleration and deceleration maneuvers at traffic signals results in significant fuel and energy consumption levels. Green light optimal speed advisory systems require reliable estimates of signal switching times to improve vehicle fuel efficiency. Obtaining these estimates is difficult for actuated signals where the length of each green indication changes to accommodate varying traffic conditions. This study details a four-step Long Short-Term Memory deep learning-based methodology that can be used to provide reasonable switching time estimates from green to red and vice versa while being robust to missing data. The four steps are data gathering, data preparation, machine learning model tuning, and model testing and evaluation. The input to the models included controller logic, signal timing parameters, time of day, traffic state from detectors, vehicle actuation data, and pedestrian actuation data. The methodology is applied and evaluated on data from an intersection in Northern Virginia. A comparative analysis is conducted between different loss functions including the mean squared error, mean absolute error, and mean relative error used in LSTM and a new loss function is proposed. The results show that while the proposed loss function outperforms conventional loss functions in terms of overall absolute error values, the choice of the loss function is dependent on the prediction horizon. In particular, the proposed loss function is outperformed by the mean relative error for very short prediction horizons and mean squared error for very long prediction horizons.

## INTRODUCTION
Consistent stop and go operation of vehicles at signalized intersections pose a significant operational challenge. From a vehicle standpoint, consistent stop and go at traffic maneuvers at signals leads to more acceleration and deceleration resulting in higher fuel consumption and lower energy efficiency. From a network perspective, consistent stop and go induces shockwaves propagating through the network that reduce roadway capacity (*1*). This leads to significant delays incurred by the network users as well as safety concerns due to aggressive acceleration and deceleration and uncertainty associated with traffic signals (*2*).

Alleviating this problem has two main sides. The first is optimizing the logic and adaptability of the traffic signal controllers to introduce operations that can minimize the number of stops. This can allow the progression of traffic flow across the network. This problem has been tackled by multiple researchers, but it is a complex task and involves significant investments by the infrastructure operators. The other side is controlling the vehicle speeds to allow them to progress through signals when they are green and thus reduce the probability of having to stop. A lot of previous work investigated the Green light optimal speed advisory (GLOSA) and different control systems to minimize stop and go operation and maximize fuel efficiency (*3; 4*). The underlying assumption for these systems, however, is that there is a reliable estimate of the switching times of upcoming traffic signals. Moreover, having a reliable estimate of the signal switching time can provide the drivers less uncertainty alleviating the issues of dilemma zone and improving safety at intersections. It is also crucial for automated and connected vehicles to make more informed decisions when approaching intersections for more stable and fuel-efficient operation.

Optimizing traffic signal controller logic and providing drivers with optimal speed advice are two conflicting objectives. The reason behind that is that the more adaptive a traffic control system is to changes in traffic conditions, the less predictable it is in terms of switching times. In other words, actuated traffic signals would provide more time to a given movement if the detectors are being actuated. Therefore, deducing the amount of green time left for a particular movement would entail knowledge of the exact pattern at which detectors will be actuated over time. These actuations are very stochastic in nature and are difficult to predict making estimation of the remaining time for a signal to switch a very challenging process.

## Problem Statement

Existing traffic signal controllers are very flexible in terms of accommodating different road users and allocating varying times to different movements. This high flexibility gives rise to very low predictability of the controller where the more flexible the controller is to accommodate different traffic conditions, the less predictable its timing is. The complexity of predicting the traffic signal state can be attributed to two factors. The first is the controller logic where the D4 controller logic is highly adaptable and allows for different ways to adapt to the incoming traffic. Examples of this adaptability include options like having floating green times that can be allocated to different movements, allowing all settings to vary according to time of day, allowing locked calls for vehicles or pedestrians to be placed on a certain phase and allowing reserving left turning vehicles that have not been serviced meaning that one cycle can have two left turn phases, if needed. While some of these features are not still used, they make the traffic signal highly unpredictable and increase the complexity of the prediction task. The second factor is the highly stochastic nature of traffic and pedestrian arrivals. Several studies have attempted to predict the effect of traffic arrivals on signal timing using low frequency probe vehicle data, GPS trajectory big data and data from upstream intersections combined with platoon dispersion modelling (*5-7*). Their results show the highly stochastic nature of traffic arrivals.

The problem lies in the lack of a holistic approach to predict the dynamic traffic signal switch times including both time to green (TTG) and time to red (TTR) with reasonable accuracy, especially in areas of high traffic demand levels. This leaves drivers less aware of when the traffic signal indication might change leading to significant fuel consumption and energy losses during acceleration and deceleration, as well as reduced throughput and less efficient operations. To be effective, this approach should consider all the different users of the transportation network as well as the underlying traffic conditions that affect the signal timing.

### Proposed Approach

This research proposes the use of Long-short term memory (LSTM) recurrent neural networks as an approach for the prediction of the signal switching times. This approach allows for not only including all the data relevant to the prediction but also recognizes the temporal dependencies between the data elements at different time steps. This is allowed by the special building block of the LSTM network known as the LSTM cell that models temporal dependency amongst variables. This feature of capturing the temporal dependencies made LSTM lend itself to areas such as language modelling, speech recognition and stock market price prediction (*8-10*). LSTM networks have also been used in Transportation applications. They have also been applied to areas with strong reliance on temporal trends such as roadway link travel time prediction, traffic flow prediction as well as accident risk and severity prediction (*11-15*).

The key idea is to recognize the importance of temporal dependencies amongst signal states at different times, as well as the temporally-dependent nature of most data used in predicting the signal switching times. This includes traffic volumes, speed, traffic arrivals and pedestrian arrivals data. These parameters not only depend on the time of day but also can show distinct trends in the very short term. For example, if the traffic volume on one of the roads is increasing over the past few cycles, it might have a higher probability of increasing in the current cycle.

### Objectives and Contribution

This research has three main objectives:

- Outline a detailed practical framework for use of LSTM in traffic signal switching time prediction.
- Test the framework on signal switching time prediction using field data.
- Compare the effect of using different loss functions on the training process.
- Propose a novel loss function that enables the predictive model to generalize well across various time horizons.

To the best of our knowledge, this research is the first to implement LSTM for actuated coordinated traffic signal switching time prediction. Unlike most studies that either focus on time to green (TTG) switching time or time to red (TTR) switching time, our proposed approach utilizes a single model for predicting both TTG and TTR switching times. This model provides a holistic approach which considers all the different users of the transportation network including the effect of pedestrian traffic. The effect of pedestrian traffic has not yet been included in any of the existing models for predicting actuated signal switching times. This can be mainly attributed to the fact that data regarding pedestrian actuations from push buttons and pedestrian phase timing and exit modes was not readily available. Finally, the modelling approach described in this paper is highly data-driven and can be extended from its current reliance on historical data to a live implementation that continuously utilizes data to improve predictions.

## METHODOLOGY

To be able to provide predictions for the traffic signal switching times, a four-step research methodology was undertaken. The first step was data collection which entailed gathering the data broadcasted on the website every second and saving it to the database. The second step was data validation and preparation which included formatting the data to generate 120 second sequences of data to be used by the LSTM Network for prediction. This involved handling missing data due to connection losses and server downtime as well as extracting the relevant data features to enhance the prediction. The third step was designing, coding, and tuning the LSTM Neural networks to improve the prediction accuracy using cross validation. The final step was validating testing the trained network on out of sample data.

## Data Description

Data was collected from Gallows road, which is part of state route 650 in Fairfax County in Northern Virginia. Gallows road is a major commuter road extending connecting between Tysons Corner, a census-designated place which is a major shopping destination in Fairfax county and Annandale, a census-designated place with multiple residential and commercial areas. The roadway has a high traffic volume where the average daily traffic is 39000 users and the average annual weekday traffic is 42000 users. The land use of areas surrounding Gallows Road is mostly commercial despite the area having a few residential developments.

Data gathering scripts have been deployed to obtain data from the intersection between Gallows Road and Gatehouse Drive (**Figure 1**). A database of historical data was created including 83 days of data for each of the intersections. The obtained data spans over the period from July 2019 till December 2019. The data was obtained from Virginia Smarterroads online portal. The online portal broadcasts data every one second which is very descriptive of the traffic state.

Data includes a detailed state of the signal including its cycle length, current phase and status and exit mode for each phase. Included in our data is a detailed description of the controller settings including signal timing plan ID, cycle length and offset from upstream signal. The very detailed and fine-grained nature of the available data allows the model to take into account all traffic conditions and road users when making a prediction for the state of the traffic signal. Two downsides to the data stream, however, are connection stability and latency. Due to the connection with the Smarterroads portal, about 10 percent data loss is incurred. This necessitates having a robust model that can handle the missing data without affecting the prediction. Moreover, data is broadcasted with random latency of around 2 seconds. While this latency does not affect our model utilizing historical data, it has to be taken into account if the model is extended into a live implementation. Due to the large amount of data involved, data from the intersection had to be saved as JavaScript Object Notation (JSON) files into a database indexed by the timestamp to the nearest second.

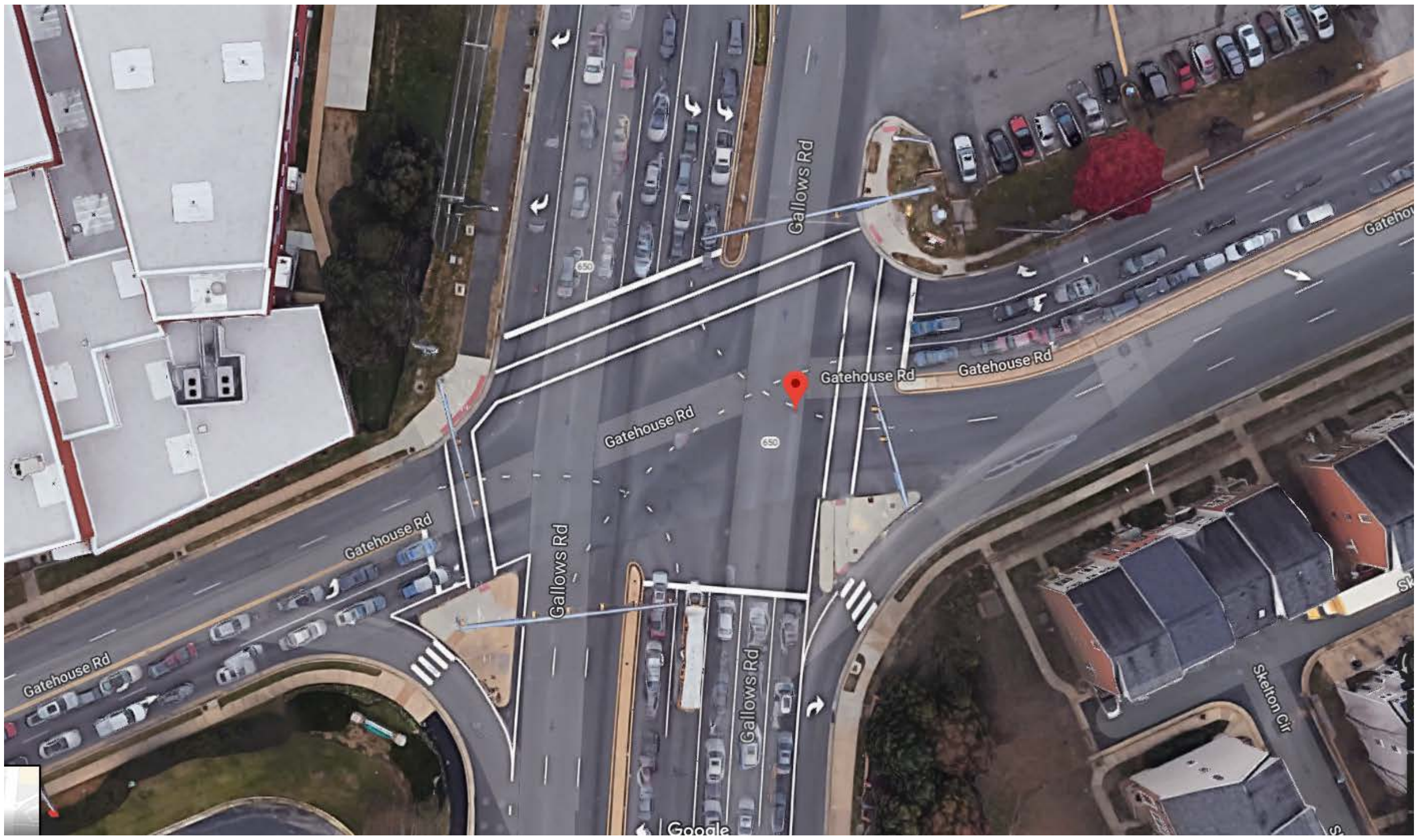

**Figure 1 Intersection layout**

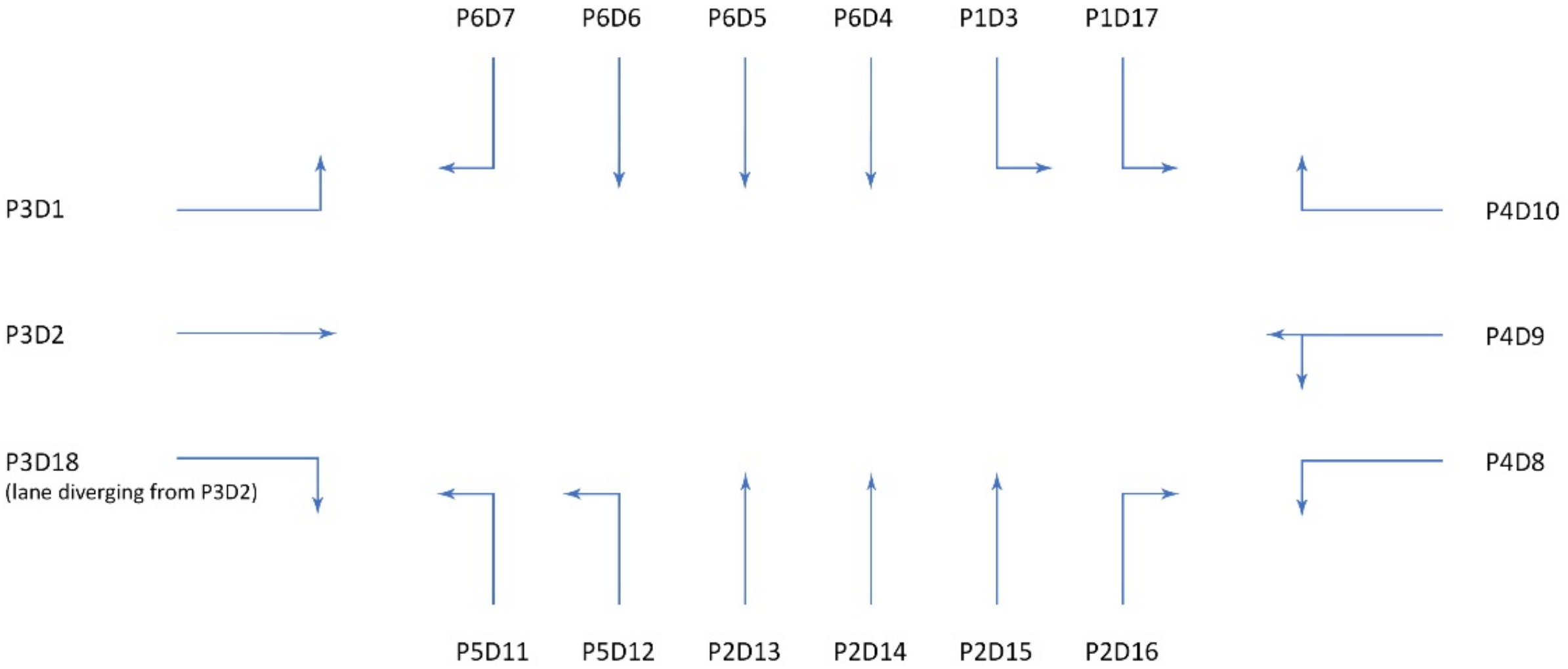

**Figure 2 lane configuration and detector placement**

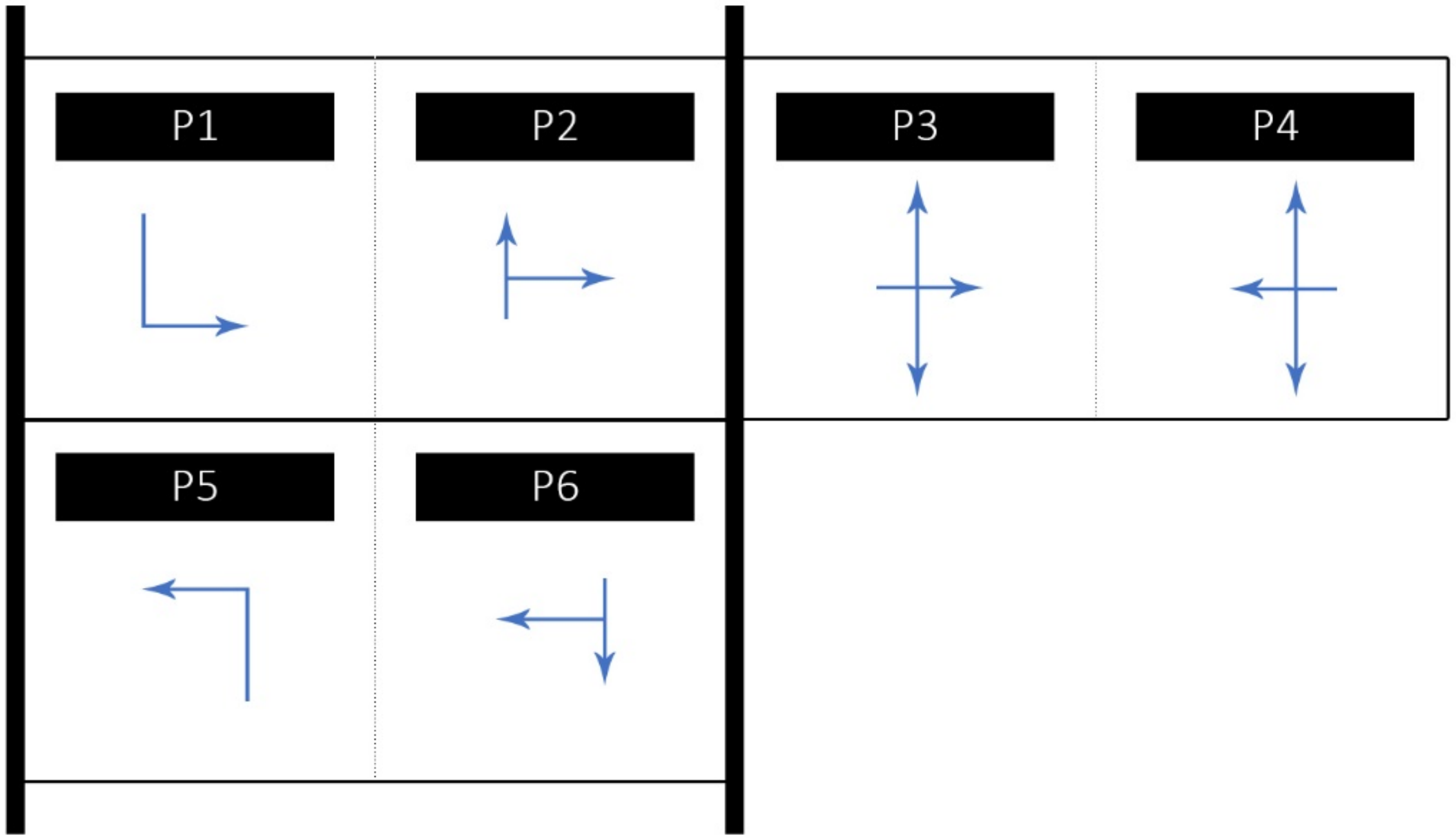

**Figure 3 Ring Barrier diagram for National Electrical Manufacturing Association (NEMA) phasing**

### *Intersection Geometry and Signal Phasing*

As shown in Figures 1 and 2, the intersection has four approaches. The northbound and southbound approaches each have 6 lanes. The eastbound approach has two lanes with a right turn lane diverging from the right lane and the westbound approach has three lanes. Figure 2 shows the detailed lane configuration as well as the placement of 18 different detectors where there is a detector for each lane. Each detector provides data that includes traffic volumes, speeds, and occupancies as well as vehicle actuations measured by detectors for each separate lane.

The data collected all spanned from 6 am till 10 pm where the signal was operated in actuated coordinated phase. This means that while the overall cycle length was predefined, there was what is called "floating green time" which could be allocated between the different signal phases. The overall cycle length was changing based on a predefined time schedule according to the time of day and day of the week.

Figure 3 shows the NEMA phasing ring barrier diagram. Phases 1, 2, 3 and 4 are all on ring 1 whereas phases 5 and 6 are on ring 2. The barrier (indicated by thick lines) separate the movements in the north south direction from the movements in the east west direction. The signal timing plan starts with the left turns for the north south directions followed by the through and right traffic. Overlap is allowed to occur between phases 1 and 6 and between phases 2 and 5. For the east-west direction, split phasing is employed where the eastbound approach is allowed to discharge all through, left and right movements followed by the westbound approach discharging all its movements. Three pedestrian phases occur simultaneously with vehicle phases 2, 4 and 6. The timing of these phases walk and do not walk as well as the pedestrian actuations/ calls and the pedestrian signal status are also reported as part of the input data. Not all these phases have to be served for each cycle, but phases can be skipped in case there is no actuations occurring for the phase which adds to the complexity of the signal state prediction.

## Data Preparation

Once all historical data was gathered into the database, this step was concerned with preparing the data to be used in the machine learning step. The data was queried from the database and converted to input files for the LSTM networks. Converting JSON data files into LSTM inputs was a multistep process that involved several data manipulation steps.

### *Creating Data frame from JSON Files*

JSON includes data in a hierarchical nested structure of variables. For example, the timing data, would be separate from the detector or pedestrian data. The JSON file was a hierarchy of data structures which included different data. For each second of data, there was a JSON file that included all the data for this second in a hierarchical structure. The first step involved extracting all the relevant data from the JSON files and converting the hierarchical tree like structure to a flat structure such that each second can be an entry in a table like structure or a data frame. The second step was filtering the data since the JSON file format was unified for different intersection and therefore had some redundant, duplicated, or irrelevant data for our intersection which was omitted. It should be noted that some variables were present in some days but marked as missing in other days. Therefore, to be able to unify the data frame structure, all data had to be examined first before deciding on what variables were to be included. If a certain day had no value for the variable it had to be marked as missing.

### *Preprocessing Numerical and Categorical Variables*

At this point data was converted from a large number of JSON files with one file per second to a more concise set of table-like data frames with one file for each day of data. The third step was to recode categorical variables into dummy variables. This involved recoding variables like "signal state" or "exit mode" or "detector actuation" into n-1 dummy binary variables where n is the number of possible states of the categorical variable. The fourth step was normalization of numerical variables to be between zero and one. This was achieved by subtracting the minimum value of the variable and dividing by the difference between the maximum and the minimum values. The maximum and minimum values for each variable were obtained from sampling several days of data throughout the study period and obtaining the absolute maximum and minimum values from these days. These maximum and minimum values were not only used for the training data, but also the same values were used for the validation and testing data. The reason this is important is that in a field implementation, the maximum and minimum values for the prediction will not be known. Therefore, the maximum and minimum values for the training data will have to be used for normalizing the variables of the field data to obtain the prediction. The only numerical variable which was handled differently was the time of day which is a cyclical variable that was coded into two variables which are $\sin(2\pi t/n)$ and $\cos(2\pi t/n)$ where t is the time of day in seconds and n is the total number of seconds per day. This mapped the time of day as a point on a circle which is common practice for deep learning data preparation. After this process each second of data was comprised of 187 variables which feed into the prediction.

### *Re-indexing Timesteps and Output Variable Computation*

At this stage, all variables are in the correct format, however, some of the seconds in each day are missing and some are duplicated due to connection. This is rectified by re-indexing all the tables by using a unique id which is chosen to be Unix time. Unix time is the number of seconds elapsed since 1 January 1970. By using the Unix time as an index, all missing intermediate data could be filled in as missing data. The missing data was given a unique value -1 as none of the variables had values below zero. This step is essential as the input to the LSTM network has to be time series equally spaced across time. The time step between subsequent data points had to be set to be exactly one second even if some of the data points in the time series are missing.

The next step was computing the output variable which is the time until the signal state changes from green to red or vice versa. The data included the state of the traffic signal at each time step but not the time until it changes state so this involved looking ahead a time horizon of 200 seconds and defining the time until the signal state changes. This was done for each of the 6 phases for the signal. It is one of the most computationally demanding data preparation tasks as it involves iterating up to 200 seconds in the future for every second where the time remaining cannot be deduced from the previous timestep time remaining. In case the time remaining cannot be deduced from the next 200 data points either due to missing data or due to phase skips prolonging the switching time beyond 200 seconds, the switching time is marked as missing and is not used for training the model. The output variable is then normalized by dividing by 200 seconds (the maximum value where the minimum value is zero).

***Generation of Sequences and Batch Processing***

Once all data was arranged with consecutive time steps in seconds, the next step was generating input for the LSTM network. LSTM network was trained in batches where each batch was a three-dimensional data structure containing 1000 sequences and 1000 predictions. Each sequence was a two-dimensional data structure or a matrix of values. The matrix consisted of vertically stacked data points describing all the input variables for the duration of the past two minutes until the second where prediction is taking place. Each time step in this sequence contains the 187 input variables obtained from the JSON files. For each of these sequence matrices, the prediction is a 6-element vector which contains the time left for each of the 6 signal phases to change. The data training data included a total of 2,164,000 sequences and their corresponding predictions making 2164 batches of training data. Handling these sequences was very memory intensive as the sequences were 387 gigabytes of data and therefore required very memory efficient operation to be able to shuffle all sequences and load them to the model in varying order.

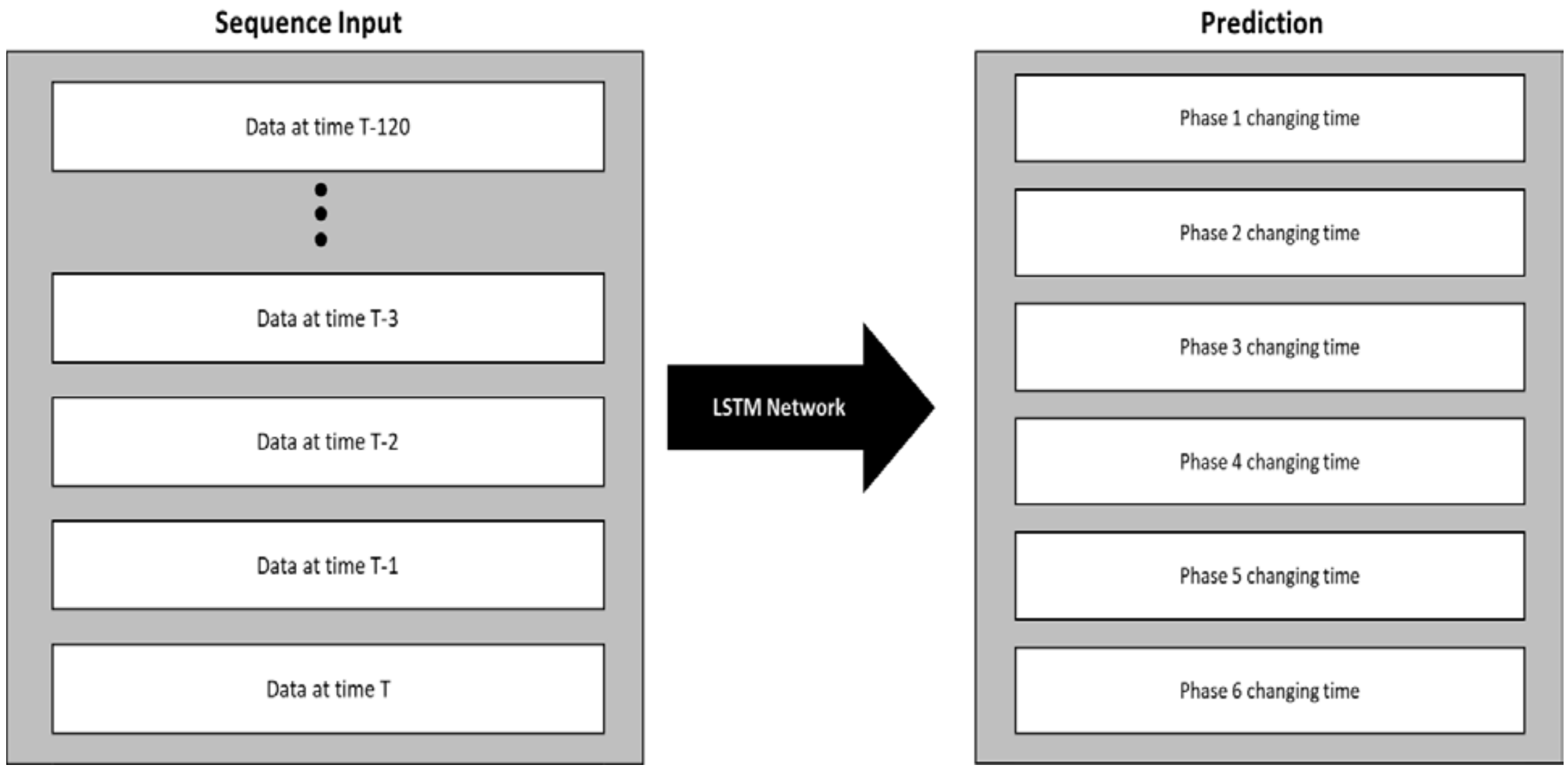

**Figure 4 Sequences and Prediction Data Structures**

## Machine Learning

### *LSTM Networks*

Recurrent neural networks (RNN) were introduced as a deep learning framework which includes temporal dependencies between temporal sequences of hidden layers. Recurrent neural networks, however, did not perform very well training on data with long term temporal dependencies. It was critiqued for the issue of vanishing and exploding gradients while training (*16*).

LSTM model architecture was proposed as an alternative for typical RNN architecture more capable of storage and access of long-term temporal dependencies. The architecture, originally proposed in 1997, has been continuously improved and adapted into modern research paradigms (*17*). The architecture used in this paper is similar to that used by Graves in 2013 and depicted by the following equations(*18*).

$$i_t = \sigma(W_{xi}x_t + W_{hi}h_{t-1} + W_{ci}c_{t-1} + b_i) \quad (1)$$

$$f_t = \sigma\left(W_{xf}x_t + W_{hf}h_{t-1} + W_{cf}c_{t-1} + b_f\right) \quad (2)$$

$$c_t = f_t c_{t-1} + i_t \, tanh(W_{xc}x_t + W_{hc}h_{t-1} + b_c) \quad (3)$$

$$o_t = \sigma(W_{xo}x_t + W_{ho}h_{t-1} + W_{co}c_{t-1} + b_o) \quad (4)$$

$$h_t = o_t \, tanh(c_t) \quad (5)$$

These five equations describe the LSTM cell basic architecture with input $x_t$ for timestep t. i, o, and f describes the input, output and forget gates, respectively. c refers to the inner cell which stores the information. h refers to the hidden state. The given architecture provides 11 sets of weights and 4 biases connecting between the different gates and the cell. Furthermore, the cell values and hidden states from previous time steps are all fed into the cell, input, output and forget gates allowing the neural network to adjust the weights and to store only the meaningful temporal dependencies between the different time steps.

### *Model Formulation*

The LSTM neural networks were built using TensorFlow Keras, which is a high-level machine learning package in Python. All the tested models consisted of 5 fully connected layers (**Figure 5**). The input layer had 187 nodes to process all data inputs. This was followed by an LSTM layer with N number of LSTM units where N is a value which was experimented with to find the best model. This was followed by a dense fully connected layer with rectified linear unit (ReLu) activation with N number of nodes. The model was set up as a regression output model where for each phase, the time remaining for the phase to change could be any value between zero and 200. This means that the output prediction was to be a linear combination of the outputs of the dense fully connected layer with ReLu activation. This means that the output layer was to have 6 nodes with linear activation. This allows the network to output the values of the time remaining for the state to change from red to green or vice versa for in each of the 6 signal phases. This means that the output is continuous variables which have then to be rescaled to 200 seconds and then approximated to the nearest second. The models all utilize Adaptive Moment Estimation (Adam) optimizer and different loss functions are experimented with including mean squared error, mean absolute error and mean absolute percentage error. The different models obtained are compared against one another and prediction errors in the short term and long term of different models are used. More details of the different models used, and their performance are discussed in the results and analysis section.

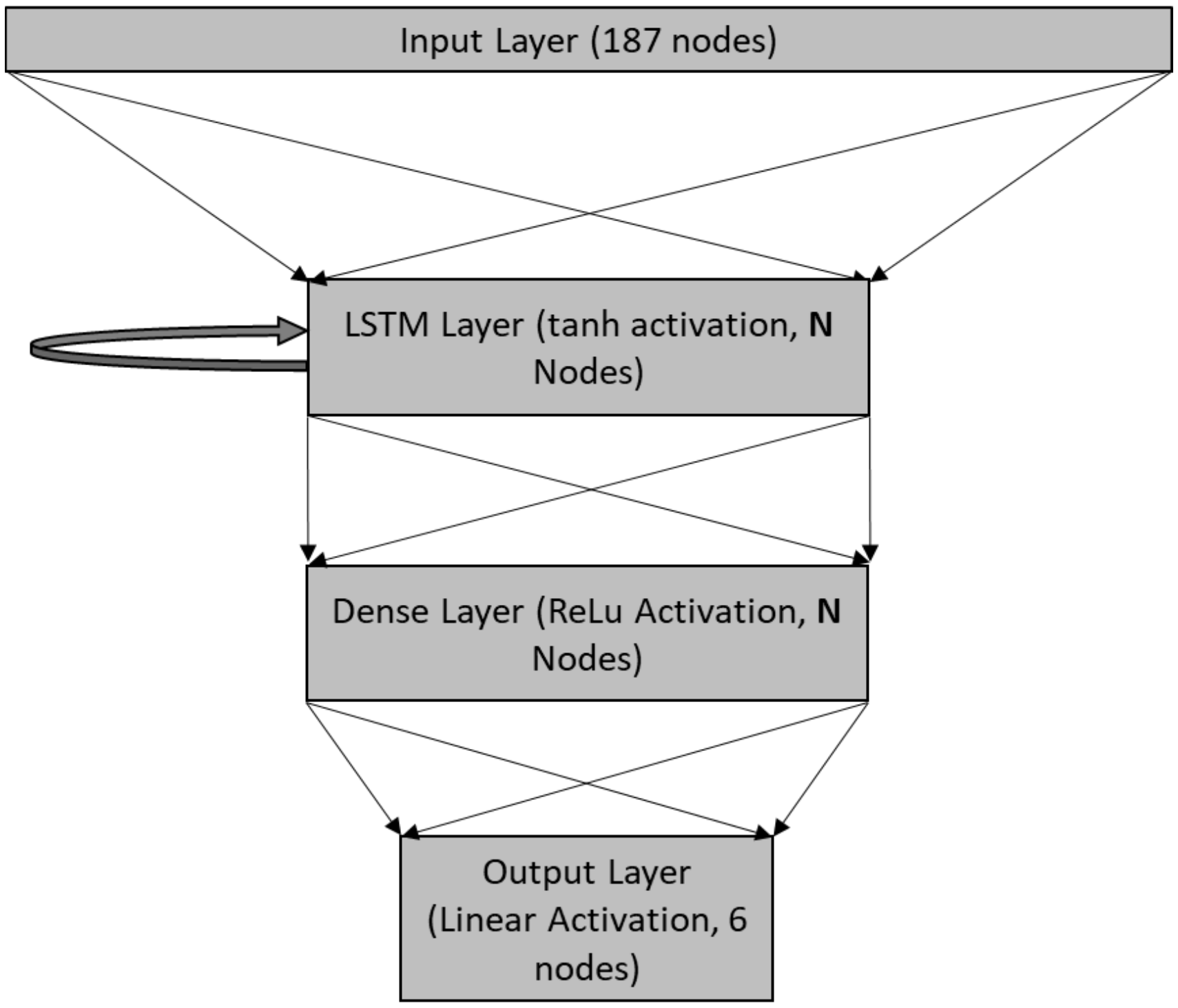

**Figure 5 LSTM Network Architecture Used**

# RESULTS AND ANALYSIS

## Data Split

Cross validation was used for every model to prevent overfitting. The data was divided to three subsets. Training data is used to train the model. Validation data is used to check the performance of the model regularly during training to avoid overfitting and tune hyperparameters. Testing data is used to evaluate the performance of the finalized model. Similar to what would happen to a field implementation, the model was trained using 81 days of data between the 26$^{th}$ of July 2019 and the 10$^{th}$ of December 2019. The validation data used for tuning the hyperparameters of the models was used as a single day of data which is the 14$^{th}$ of December 2020. The testing data used for evaluating the performance of the finalized models was the 16$^{th}$ of December 2020. The validation and testing data were chosen to be for a date in the future outside the bounds of the training data. The reason for that is that the key interest lies in the ability of the models to generalize to future data rather than generalizing to historic data the model has not been exposed to. Furthermore, the use of a single day for validation and a single day for testing was done as a single day worth of data contains a representative amount of data to assess the entire model performance. The fact that each second of data is considered a data point means that a single day of data contains enough data points to allow a model to be thoroughly assessed.

## Exploratory Model Architecture Experiments

Two key model hyper-parameters were to be identified for the best model. The first is the value of N which is the number of neurons and the second is the learning rate to be used for Adam optimizer. An exploratory model fitting experiment was done on 37 days from July 26, 2019 till October 3, 2019. The data from October 4, 2019 was used for validation. The 37 days included 1.1 million data points. Due to the large computational time associated with the sequence generation, loading data and model training, these models were limited to going over the data twice while monitoring the loss function and the performance on the validation data every time the model finishes processing 250,000 datapoints. A grid search was utilized to explore the hyper-parameter space for different combinations of learning rates and number of neurons per layer.

The choice of learning rates was on a logarithmic scale. For the choice of number of neurons, the value 187 is equal to the size of the input. It is then reduced twice by a factor of 4 to 47 and 12 respectively and then the size of the output layer 6 is used. Mean relative error was used as a loss function and performance metric to compare different models in this case for two reasons. The first is that the absolute value of the error is less important the more the prediction horizon so for example an error of 5 seconds in a 10 second horizon is much more than an error of 5 seconds in a 50 second horizon. The second is that previous papers which employed LSTM to travel time and traffic flow forecasting which are the most relevant areas to our application domain utilized it as a key performance metric (*19*-*21*). Its performance against other loss functions would be tested in the following section.

**TABLE 1 Mean absolute percentage error performance on Validation Data**

| Learning Rate | Number of Neurons | | | |
|---|---|---|---|---|
| | 6 | 12 | 47 | 187 |
| 1*10^-2 | 22.61 | 21.03 | 17.57 | 20.00 |
| 1*10^-3 | 28.56 | 33.00 | 20.18 | 27.36 |
| 1*10^-4 | 73.38 | 52.51 | 41.73 | 39.16 |
| 1*10^-5 | 223.27 | 269.53 | 196.74 | 108.81 |
| 1*10^-6 | 364.86 | 425.77 | 382.64 | 369.63 |

Table 1 shows the minimum reached mean absolute percentage error for different combinations of number of neurons per layer N and learning rate. One key insight that can be drawn from this analysis is that with the complexity and large computational time involved, choosing the correct learning rate is a key determinant of the quality of the model reached where the variation of the mean absolute percentage error over the varying learning rates is much larger than over varying number of neurons. Another observation is that the 47 neuron is better performing at both the 0.01 and 0.001 learning rates. Accordingly, the number of neurons would be set to 47 for further analysis.

## Performance of Different Loss Functions

### *Model Description*

The maximum prediction horizon for the models used is 200 seconds. This is 3 minutes and 20 seconds in the future and there are multiple uncertainties affecting predictions further ahead in the future. This property of having more certainty in predictions which are in the near future and less certainty in predictions in the distant future is common many problem domains and predicting traffic signal switching times is one of these domains. This necessitates the choice of a proper loss function to reflect this property to be able to properly balance the prediction over short term and long-term prediction horizons. Three commonly used loss functions are examined which are mean squared error, mean absolute error and mean absolute percentage error which is sometimes referred to as relative error. Another loss function is proposed by the authors which is a modified version of the mean squared error which can be expressed as:

$$L = \left(y_{pred} - y_{true}\right)^2 * (1 - y_{true})^2 \quad (6)$$

This function modified the mean squared error function by scaling it to be smaller as the true value of why is greater. It should be noted in this case that the true value of y has to be scaled between zero and one. This both uses the squared value of the error as well as having the property of scaling down the loss as the ground truth value of the time remaining is further ahead in the future similar to relative error. However, it provides more flexibility than relative error as the value of relative error increases indefinitely as the ground truth approaches zero whereas the proposed loss function would be approaching the sum of squared error as the ground truth value approaches zero.

All four loss functions are used to fit the entire dataset with the model discussed in Figure 5 with N=47. According to the findings from the exploratory networks, the learning rate was set as 0.01. Despite ADAM optimizer having its own adaptive learning rates based in momentum, an additional learning rate decay is added which multiplies the learning rate by 0.3 for every epoch where there is no improvement in the validation loss. This allows the learning rate to go down an order of magnitude for 2 epochs with no improvement. All 2,164,000 data points are fitted using each of the four loss functions discussed over 10 epochs allowing the model to assess the performance on validation data each epoch. For every loss function, the model with the best loss over validation data is selected and then the best model over the 10 epochs for each loss function is compared against those of other loss functions.

***Model Performance Comparison***

After fitting the models and deciding the best model for each loss function in terms of validation loss, the final models are evaluated by applying them to the testing data. The testing data contains 29578 sequences and predictions. In terms of overall model performance over the 200 second prediction horizon, the proposed model has the least absolute errors (**Figure 6**). Mean squared error and mean absolute error functions follow with mean square error having higher probability to be within 5 seconds from the ground truth but absolute error having higher probability to be within any error range bigger than 5 seconds. Mean absolute percentage error has the worst overall performance over the entire prediction horizon even though it has a higher probability than mean absolute error being up to 2 seconds from ground truth.

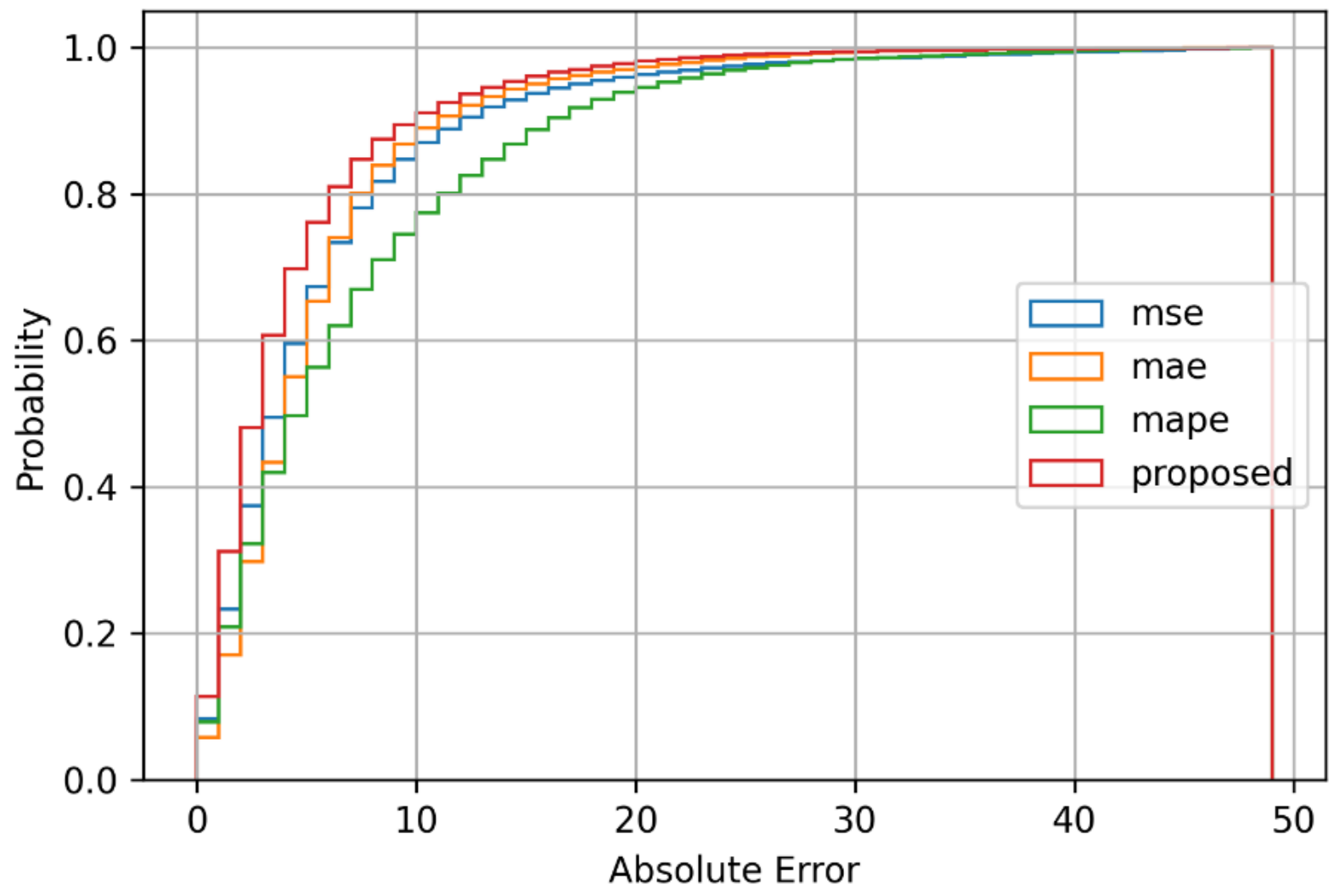

**Figure 6 comparison of the model error cumulative distribution for different loss function**

If the performance at different prediction horizon is compared between the models, the role of the loss function in optimizing the predictions becomes more evident. Both mean squared and mean absolute errors do not value how far into the future the prediction is. This means that an error of 5 seconds results into the same mean squared error and mean absolute error whether the difference is between 10s prediction and 15s truth or whether it is between 100 seconds prediction and 105 seconds truth. This results in the performance of these two models being much better in the long-term predictions. Mean squared error has the lowest error for any prediction more than 100 seconds into the future (**Figures 7-10**) (**Table 2**). Mean absolute error has slightly lower error than the mean square error in the 80 second horizon and slightly higher error after 80 seconds.

The mean absolute percentage error (MAPE), on the other hand, heavily sacrifices long term predictions for short term prediction. It has the lowest error in the 20 second horizon and outperforms mean squared error and mean absolute error up to the 60 second horizon. It can also be clearly seen that it has much less distant outliers than other loss functions for values up to 40 seconds (**Figure 8**). This, however, comes at the expense of significant reduction in longer term prediction accuracy.

The proposed model is a middle ground between the two extremes which provides an overall better fit for the data. Despite being slightly outperformed by the mean absolute percentage error for the 20 second prediction horizon, it outperforms it on every other horizon. Furthermore, it provides the lowest error for up to 100 seconds prediction horizon and is slightly outperformed by the mean squared error in the 100 second to 120 second horizon.

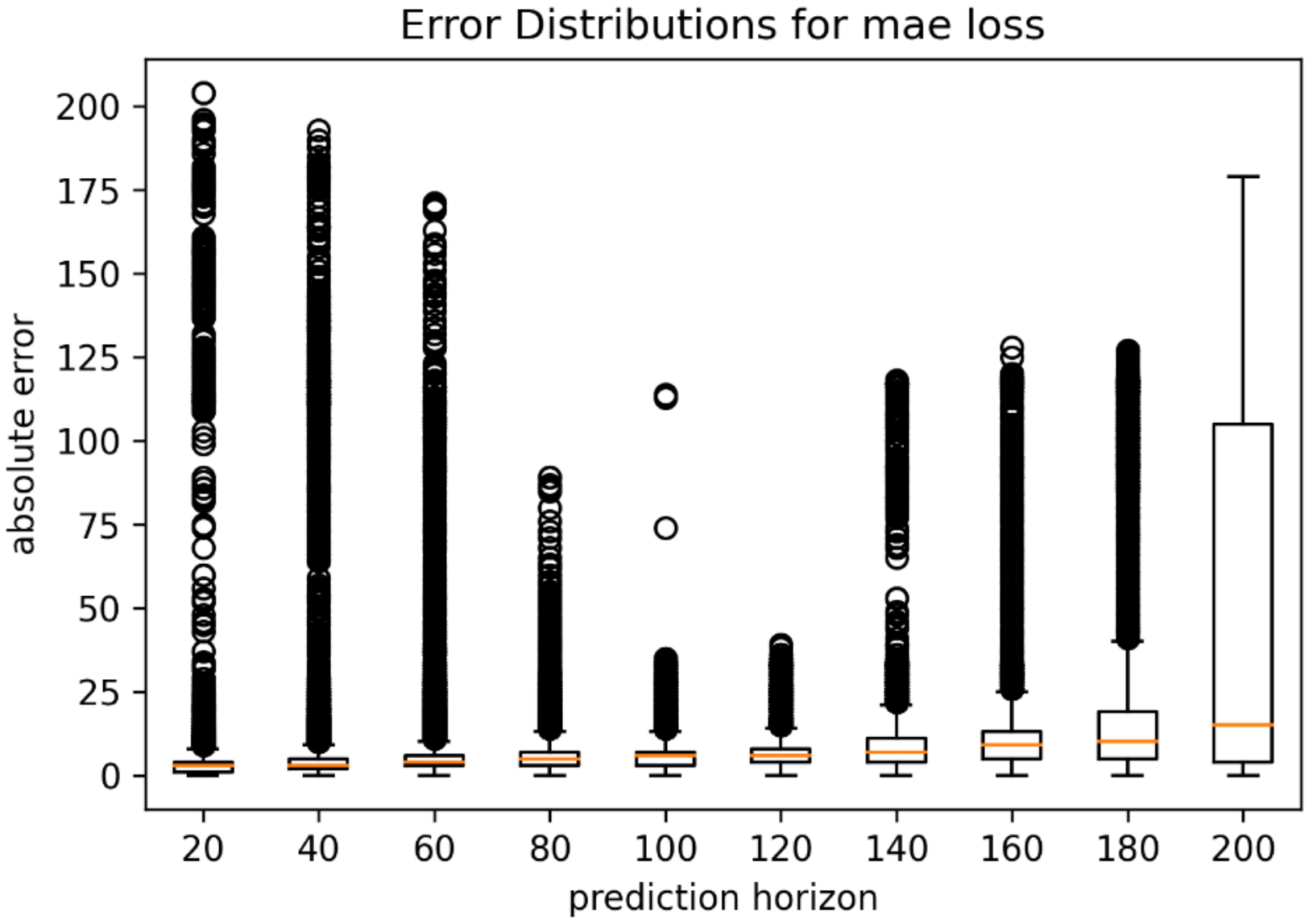

**Figure 7 Boxplots for absolute error distribution over different prediction horizons for mean absolute error function best model**

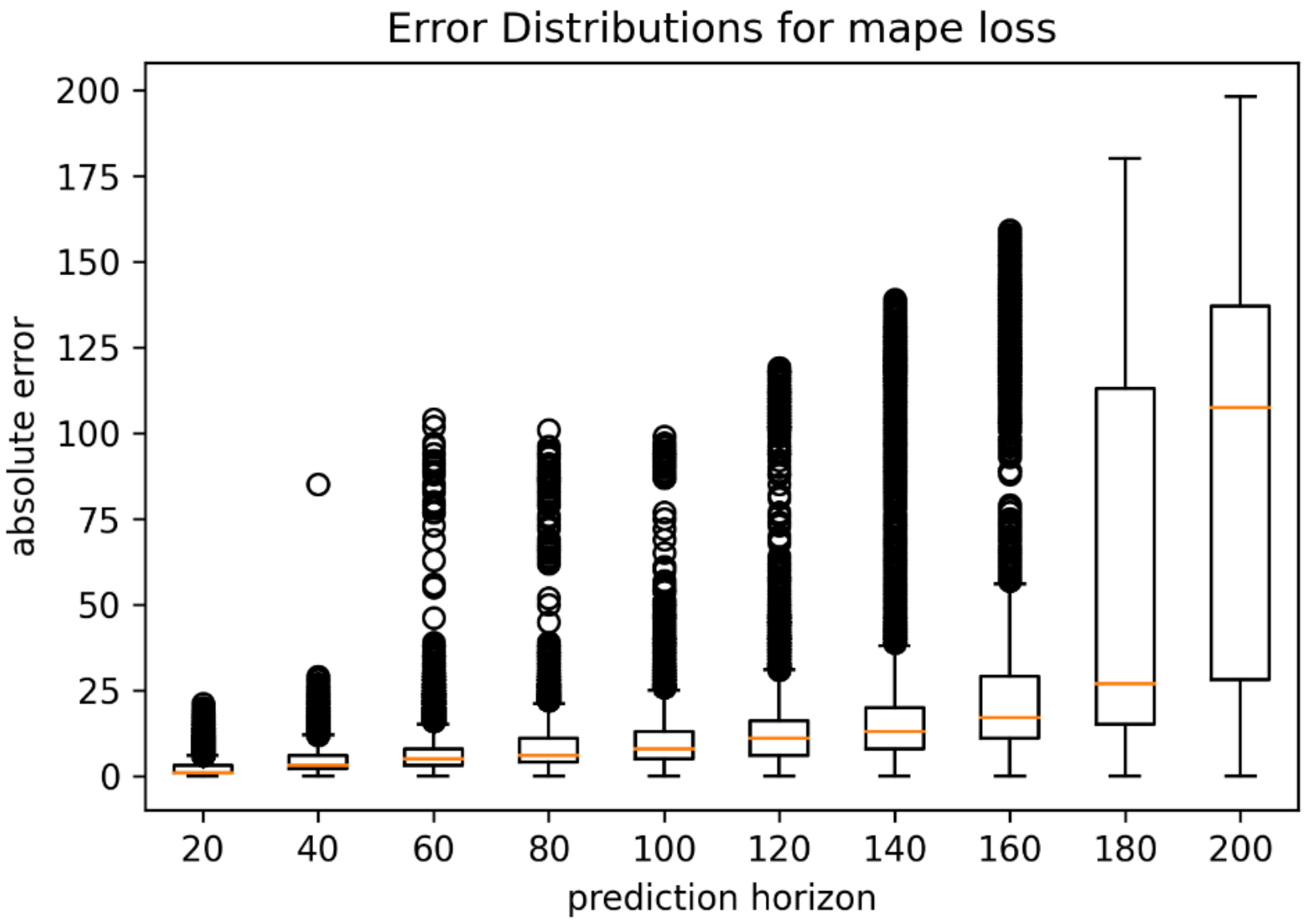

**Figure 8 Boxplots for absolute error distribution over different prediction horizons for mean absolute percentage error function best model**

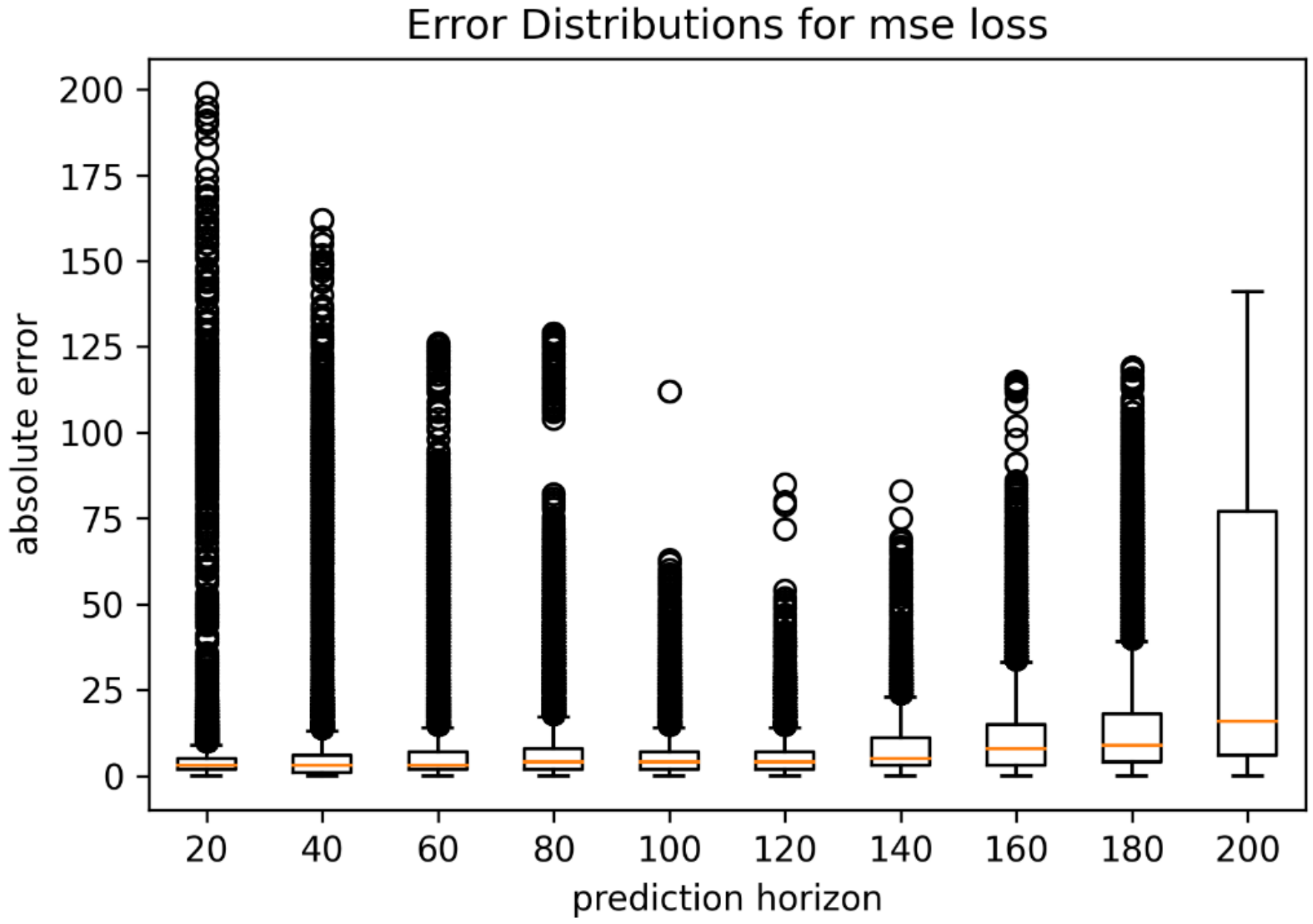

**Figure 9 Boxplots for absolute error distribution over different prediction horizons for mean squared error function best model**

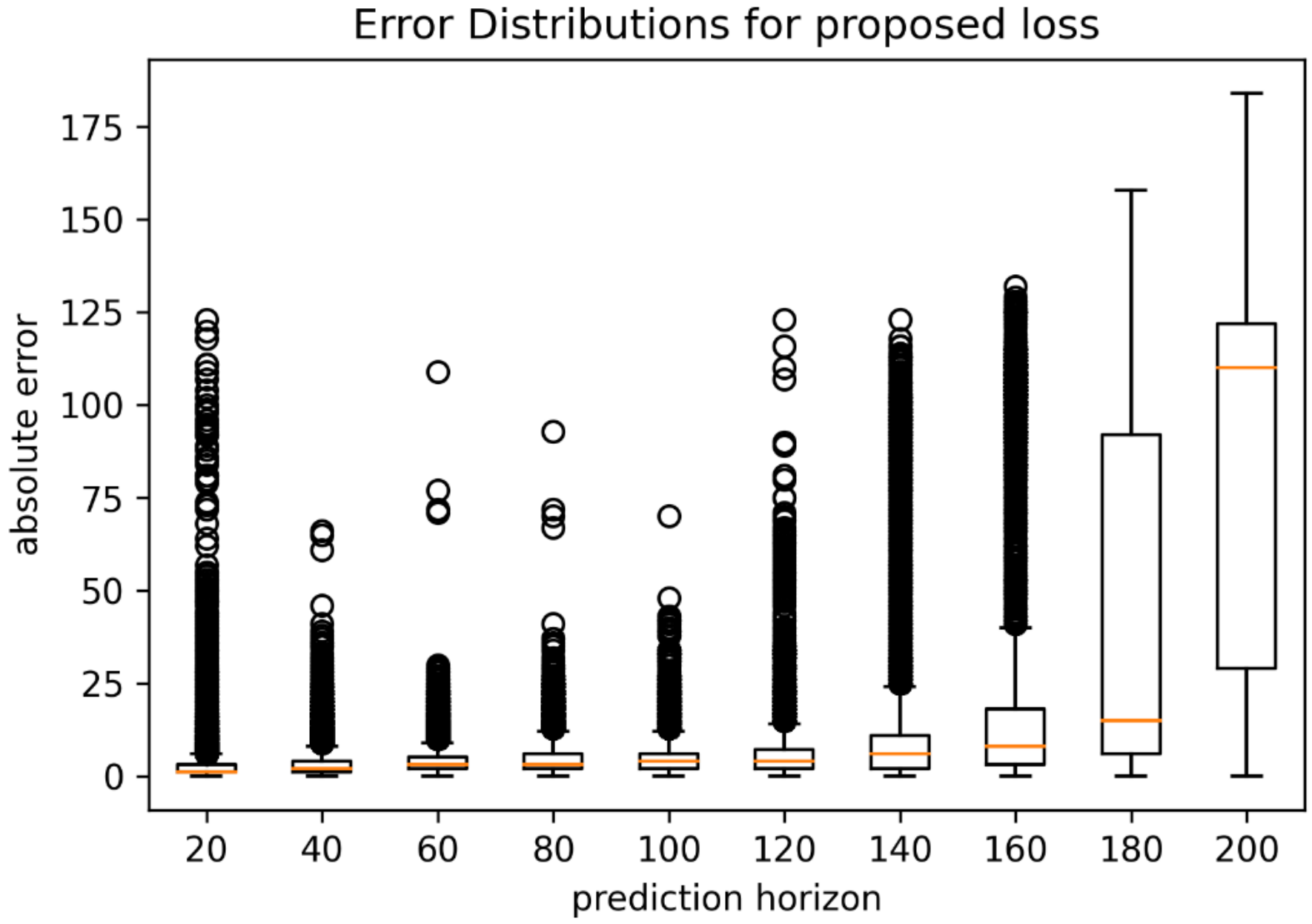

**Figure 10 Boxplot for absolute error distribution over different prediction horizons for proposed error function best model**

**TABLE 2 Mean absolute error values over different prediction horizons for the best models using each loss function comparison**

| Loss Function | Prediction Horizon (s) | | | | | | | | | |
|---|---|---|---|---|---|---|---|---|---|---|
| | 0-20 | 20-40 | 40-60 | 60-80 | 80-100 | 100-120 | 120-140 | 140-160 | 160-180 | 180-200 |
| MSE | 4.31 | 6.15 | 7.10 | 7.48 | 5.89 | 5.61 | 7.82 | 11.93 | 18.88 | 37.49 |
| MAE | 3.61 | 5.94 | 6.52 | 6.18 | 6.10 | 7.20 | 9.63 | 11.97 | 22.51 | 46.26 |
| MAPE | 1.96 | 4.20 | 6.08 | 8.09 | 10.03 | 13.46 | 26.17 | 35.49 | 52.40 | 89.45 |
| Proposed | 2.13 | 3.38 | 4.20 | 4.68 | 4.71 | 5.74 | 13.27 | 23.22 | 40.59 | 89.35 |

## CONCLUSIONS

The use for LSTM neural networks allowed for a holistic data driven modeling framework which takes into account controller logic, time of day, vehicle actuation data and pedestrian actuation data. The presented framework is robust to missing data and provides useful insights into signal switching times whether it is time to green or time to red.

The study presented a step by step detailed methodology for data gathering, data preparation, training and tuning the LSTM models and testing and comparing different model architectures. The model was applied to the four-way traffic intersection between Gallows Road and Gatehouse Drive and provided predictions for up to 200 seconds in the future relying on the past two minutes of data. A comparison between different loss functions for training the LSTM network on our problem domain was conducted. In addition to the mean squared error, mean absolute error, and mean absolute percentage error, a new loss function was proposed. The overall performance of the proposed loss function was better than that of the conventional loss functions. The comparison between loss functions indicated the importance of the choice of loss function according to the desired results. For our data, **i**f the priority is very short-term prediction horizon less than 20 seconds then the mean absolute percentage error is the best option. If the priority is prediction up to 100 seconds, then the proposed model is the best option. For long term predictions more than 200 seconds in the future, the mean square error is the best option. This study highlights the importance of choice of loss function to suit the usage of the prediction.

## ACKNOWLEDGEMNTS

This effort was funded by the University Mobility and Equity Center (UMEC).

## AUTHOR CONTRIBUTIONS STATEMENT

The authors confirm contribution to the paper as follows: study conception and design: Eteifa and Rakha; data collection: Eteifa; analysis and interpretation of results: Eteifa, Rakha, and Eldardiry; draft manuscript preparation: Eteifa, Rakha, and Eldardiry. All authors reviewed the results and approved the final version of the manuscript.